%% file: name_v1.tex
\documentclass[runningheads]{llncs}
\usepackage[T1]{fontenc}
\input{src/packages}
\input{src/acronyms}
\usepackage{tikz}

\newcommand\copyrighttext{ This preprint has been accepted to the 30th International Conference on Enterprise
Design, Operations, and Computing (EDOC 2026). This version does not reflect post-submission improvements or corrections. \the \year{}\textsuperscript{\textcopyright }}

\newcommand\copyrightnotice{%
\begin{tikzpicture}[remember picture,overlay]
\node[anchor=south,yshift=20pt] at (current page.south) {\fbox{\parbox{\dimexpr0.75\textwidth-\fboxsep-\fboxrule\relax}{\copyrighttext}}};
\end{tikzpicture}%
}

\begin{document}

\title{TrustBOM: A Scalable Architecture for Confidentiality-Preserving SBOMs \\Across Organizations}

\titlerunning{TrustBOM: Confidentiality-Preserving SBOMs Across Organizations}
\author{
Van Thang Nguyen\inst{1} \and
Frederic Rupprecht\inst{1} \and
Tom Lawrence\inst{1} \and \\
Lucca Di Benedetto\inst{1} \and
S\"oren Schubert\inst{1} \and
Amor Rezgui\inst{1} \and \\
Sebastian Werner\inst{2} \and
Maria C. Borges\inst{2} \and
Stefan Tai\inst{2}
}

\institute{
Technische Universit\"at Berlin, Germany \\
\email{
\{van.thang.nguyen,
f.rupprecht,
t.lawrence,
di.benedetto,
soeren.schubert,
amor.rezgui\}@campus.tu-berlin.de} \and
Information Systems Engineering, Technische Universit\"at Berlin, Germany\\
\email{
\{sw,mb,st\}@ise.tu-berlin.de
}
}
\authorrunning{V. T. Nguyen et al.}
\maketitle    
\copyrightnotice

\begin{abstract}
\input{sections/00_abstract}

\end{abstract}

\input{sections/01_introduction}
\input{sections/02_background}
\input{sections/03_related_work}
\input{sections/04_system_design}
\input{sections/05_evaluation}
\input{sections/06_conclusion}

\bibliographystyle{splncs04}
\bibliography{references}

\end{document}

%% file: src/packages.tex
\usepackage{cite}
\usepackage{algorithmic}
\usepackage{graphicx}
\usepackage{textcomp}
\usepackage{xcolor}
\usepackage[acronym]{glossaries}
\usepackage[hidelinks]{hyperref}
\usepackage{cleveref}
\usepackage{comment}
\usepackage{xurl}
\usepackage{svg}
\usepackage[linesnumbered,ruled,vlined]{algorithm2e}

\crefname{algocf}{algorithm}{algorithms}

%% file: src/acronyms.tex
\newacronym[longplural={Software Bills of Materials}]{sbom}{SBOM}{Software Bill of Materials}
\newacronym{smt}{SMT}{Sparse Merkle Tree}
\newacronym{smts}{SMTs}{Sparse Merkle Trees}
\newacronym{zkp}{ZKP}{Zero-Knowledge Proof}
\newacronym{purl}{PURL}{Package URL}
\newacronym{slsa}{SLSA}{Supply-chain Levels for Software Artifacts}
\newacronym{zkvm}{zkVM}{Zero-Knowledge Virtual Machine}
\newacronym{zksnark}{zk-SNARK}{Zero-Knowledge Succinct Non-Interactive Argument of Knowledge}
\newacronym{zkstark}{zk-STARK}{Zero-Knowledge Scalable Transparent Argument of Knowledge}
\newacronym{ipfs}{IPFS}{InterPlanetary File System}
\newacronym{cid}{CID}{Content Identifier}
\newacronym{spdx}{SPDX}{Software Package Data Exchange}
\newacronym{ntia}{NTIA}{National Telecommunications and Information Administration}

%% file: sections/00_abstract.tex
Software Bills of Materials (SBOMs) have emerged as a key mechanism for software supply chain governance in enterprise architectures. However, their adoption across organizations remains limited due to concerns about exposing sensitive dependency information.
To address this limitation, we propose \textsc{TrustBOM}, a scalable architecture for confidentiality-preserving SBOMs integrated into enterprise CI/CD workflows. \textsc{TrustBOM} enables software providers to attest that specific vulnerabilities or restricted licenses are absent from their software without revealing the underlying dependency graph. This is achieved using zero-knowledge non-membership proofs, which are applied selectively based on consumer-defined policy constraints.
The architecture ensures that proof generation scales linearly with the number of asserted constraints rather than with the size of the SBOM, enabling efficient operation in large-scale enterprise environments. Empirical evaluation demonstrates linear performance, with an average proof generation time of 0.9 seconds per constraint on commodity hardware, indicating the feasibility of deployment in enterprise platform ecosystems.

%% file: sections/00_keywords.tex
software supply chain security \and software bill of materials \and confidential compliance \and CI/CD \and enterprise governance

%% file: sections/01_introduction.tex
\section{Introduction}\label{sec:intro}
Modern enterprise architectures are complex systems composed of many open-source components and third-party dependencies. Managing this growing software supply chain introduces several legal and security challenges. Each dependency carries its own licensing obligations, while simultaneously expanding the attack surface through vulnerabilities it may introduce.

\Glspl{sbom} have emerged as a way to tackle these challenges \cite{ntia_sbomdefinition_2019,Xia_SBOM_2023}. For software providers, they enable systematic dependency tracking; for consumers, they provide visibility into what is deployed; for regulators, they facilitate compliance auditing across the supply chain. Their relevance has grown significantly following a U.S. executive order\footnote{\scriptsize\url{https://www.nist.gov/itl/executive-order-14028-improving-nations-cybersecurity}, [Accessed: 6 Feb. 2026]} mandating SBOMs for all software sold to government agencies, and the EU Cyber Resilience Act\footnote{\scriptsize\url{https://eur-lex.europa.eu/legal-content/EN/TXT/PDF/?uri=OJ:L_202402847}, [Accessed: 6 Feb. 2026]}, which similarly recommends SBOM generation for software providers.

However, the disclosure of \glspl{sbom} for software providers introduces a critical tension between transparency and the protection of intellectual property.
Despite regulatory pressure, providers frequently limit \gls{sbom} disclosure to protect proprietary information~\cite{Xia_SBOM_2023}, as exposing a complete dependency graph may reveal sensitive algorithms or trade secrets~\cite{Xia_SBOM_2023,Xia_blockchain_2024}. Dependencies may also signal strategic intentions: cloud platform SDKs could suggest infrastructure migration plans, analytics libraries may indicate upcoming tracking features, machine learning frameworks could reveal AI capabilities in development, and internationalization packages might hint at market expansion. From a security perspective, exposed dependencies on vulnerable versions also enable attackers to systematically identify targets before patches are deployed, as demonstrated during the Log4Shell crisis (CVE-2021-44228)~\cite{Williams_ResearchDirections_2025}.
Still, consumers require assurances that the software they purchase does not violate their security and compliance requirements (i.e., does not include specific vulnerable components, legally restricted
licenses, or organizationally prohibited dependencies). Relying solely on the self-attestation of a software provider creates a trust gap~\cite{Williams_ResearchDirections_2025}. 
Recent works attempt to bridge this gap through cryptographic techniques such as blockchain-based attestation and zero-knowledge proofs (ZKPs) \cite{Xia_blockchain_2024}. This research suggests a promising direction but still leaves much of the solution space unexplored. Blockchain transactions and ZKPs introduce considerable computational and financial costs, which, if not considered carefully, can quickly become prohibitive in enterprise settings, where software is shipped frequently and dependencies are numerous. A further challenge is that any such mechanism must integrate naturally into existing enterprise platforms, rather than imposing additional operational burden on the software delivery process.

In this paper, we propose \textsc{TrustBOM}, a confidentiality-preserving SBOM architecture embedded directly into enterprise CI/CD workflows. Using cryptographic non-membership zero-knowledge proofs, software providers can attest to different consumers that specific CVEs or restricted licenses do not appear in their software, while concealing the dependency graph entirely. \textsc{TrustBOM} operates on a public blockchain and leverages an efficient Sparse Merkle Tree for SBOM representation. Proofs are applied selectively and scale linearly with the number of non-membership constraints rather than with SBOM size. The result is a cost-efficient mechanism for verifiable compliance in software supply chains, addressing key practical barriers that have so far limited SBOM use across organizations. 
This paper is based on the key insight that compliance can be verified without disclosure by shifting from data sharing to proof-based attestation.

This paper makes the following contributions:
\begin{itemize}
    \item Architecture: We design a confidentiality-preserving SBOM architecture that enables verifiable compliance across organizations without disclosing dependency graphs.
    \item Protocol: We introduce a two-phase protocol that decouples SBOM commitment from compliance verification, enabling on-demand proofs against evolving policy constraints.
    \item Scalability: We demonstrate that zero-knowledge non-membership proofs scale linearly with the number of policy constraints rather than SBOM size.
    \item Evaluation: We provide an empirical evaluation using real-world SBOM data, showing practical performance and cost feasibility in enterprise environments.
\end{itemize}

The remainder of the paper is structured to first provide technical background and related work in \Cref{sec:background,sec:related_work}. \Cref{sec:design} introduces the system architecture and threat model. \Cref{sec:implementation} describes the implementation. \Cref{sec:eval} presents the empirical evaluation, followed by conclusions in \Cref{sec:conclusion}.

%% file: sections/02_background.tex
\section{Background}\label{sec:background}

This section introduces the technical foundations underlying TrustBOM.

\subsection{Software Bill of Materials}
A Software Bill of Materials (SBOM) is a structured inventory of software components, including dependencies, versions, suppliers, and licenses~\cite{ntia2021sbom}. SBOMs are typically generated during CI/CD pipelines and serve as a basis for vulnerability detection by matching component identifiers (e.g., PURLs) against vulnerability databases such as OSV\footnote{\scriptsize\url{https://osv.dev/}, [Accessed: 6 Feb. 2026]} or the National Vulnerability Database\footnote{\scriptsize\url{https://nvd.nist.gov/}, [Accessed: 6 Feb. 2026]}. They enable transparency and compliance auditing across software supply chains.

\subsection{Zero-Knowledge Proofs}
Zero-knowledge proofs (ZKPs) enable a prover to convince a verifier of a statement’s validity without revealing additional information~\cite{doi:10.1137/0218012}. 
Modern ZKP systems \cite{8726497} and zero-knowledge virtual machines (zkVMs) such as RISC Zero\footnote{\scriptsize\url{https://dev.risczero.com/}, [Accessed: 6 Feb. 2026]} allow proofs to be generated from programs written in general-purpose languages, reducing the complexity of constructing proof circuits.

\subsection{Sparse Merkle Trees}
Sparse Merkle Trees (SMTs) are cryptographic data structures that enable efficient membership and non-membership proofs over large key spaces~\cite{10.1007/978-3-319-47560-8_13}. By mapping identifiers deterministically to tree positions, SMTs allow proofs of absence to be constructed without revealing other elements in the dataset.

%% file: sections/03_related_work.tex
\section{Related Work}\label{sec:related_work}

Existing approaches support SBOM exchange with varying levels of security and functionality. Tools such as SBOM.sh\footnote{\scriptsize\url{https://sbom.sh/}, [Accessed: 6 Feb. 2026]} and the CycloneDX BOM Exchange API\footnote{\scriptsize\url{https://github.com/CycloneDX/cyclonedx-bom-repo-server}, [Accessed: 6 Feb. 2026]} enable standardized SBOM sharing but lack mechanisms for confidentiality or verifiable claims.

Recent research explores blockchain-based SBOM management and selective disclosure. Xia et al. ~\cite{Xia_blockchain_2024} propose a credential-based architecture for SBOM sharing, identifying zero-knowledge proofs as a potential mechanism for need-to-know disclosure, but do not implement scalable proof generation. Song et al.~\cite{10936765} introduce BC-SBOM, which stores SBOM data on-chain to improve availability, at the cost of confidentiality. CredChain~\cite{9343074} enables selective disclosure using verifiable credentials, but does not support proving the absence of components.

In contrast, TrustBOM enables verifiable compliance without disclosure by using zero-knowledge non-membership proofs. Its two-phase architecture decouples SBOM commitment from verification, allowing providers to generate proofs for evolving policy constraints without revealing dependency information. To our knowledge, this is the first system to integrate non-membership proofs into enterprise CI/CD workflows with demonstrated constraint-linear scalability.

%% file: sections/04_system_design.tex
\section{TrustBOM Architecture}\label{sec:design}
This section presents the \textsc{TrustBOM} architecture, which enables confidentiality-preserving SBOMs with verifiable compliance across organizations. The design resolves the tension between protecting proprietary dependency information and providing cryptographic assurances about software composition.

\textsc{TrustBOM} follows a two-phase architecture that decouples SBOM commitment from compliance verification. This separation allows software providers to capture the software state once during the build process and subsequently generate proofs for evolving policy constraints without rebuilding artifacts.

\begin{figure}[t]
    \centering
    \includegraphics[width=0.7\linewidth]{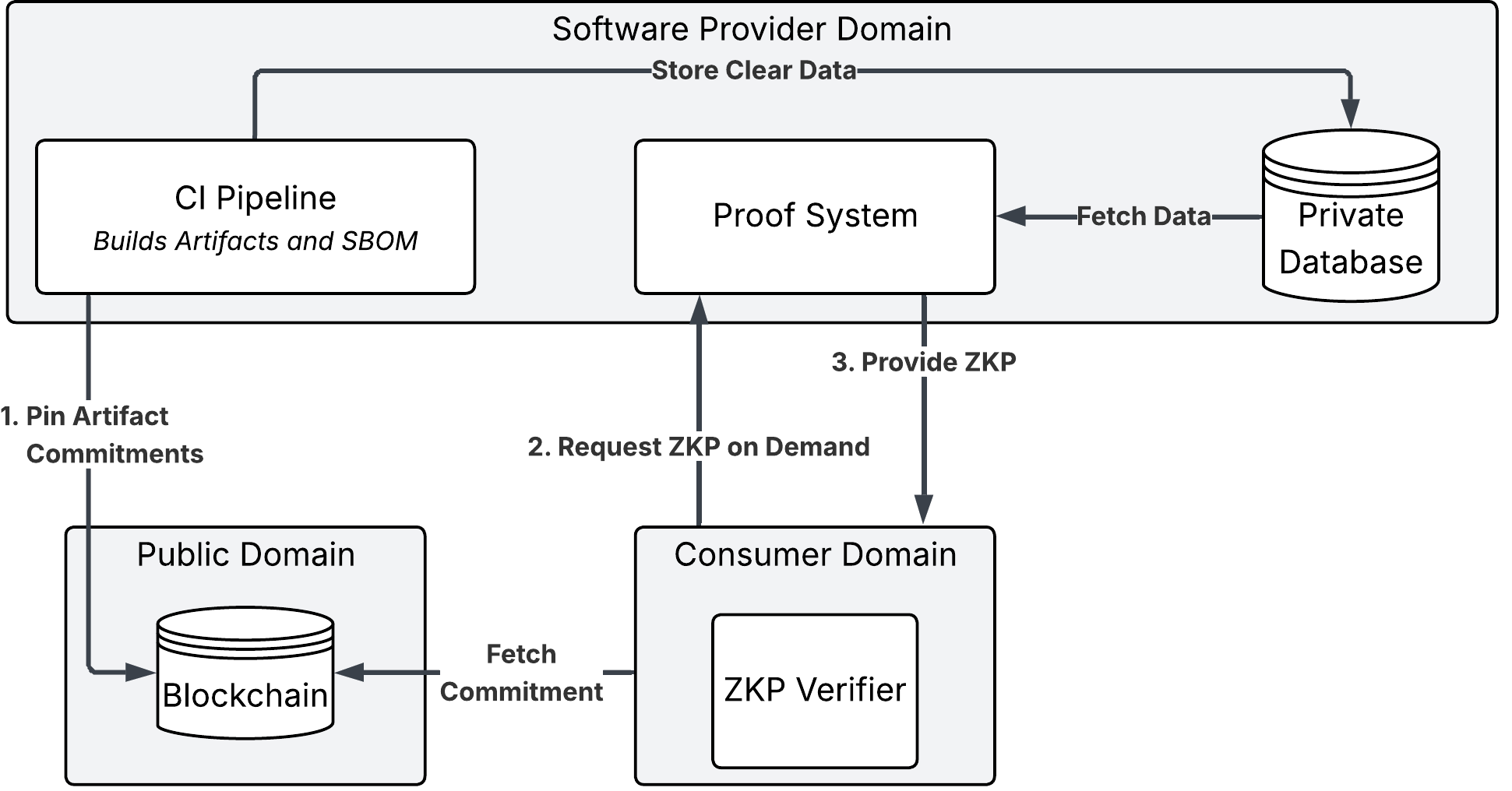}
    \caption{High-Level System Architecture}
    \label{fig:sys_overview}
\end{figure} 

\Cref{fig:sys_overview} illustrates the high-level architecture, which coordinates interactions between a software provider, a consumer, and a public blockchain. The system operates in two phases, separating software build and commitment from the generation of non-membership proofs for consumer-defined policy constraints.

In the first phase, the provider’s CI pipeline builds the software artifact, generates an SBOM, and encodes it as a Sparse Merkle Tree (SMT). The root hash of this tree is committed to a public blockchain, establishing a cryptographic commitment to the software’s dependency structure while preserving confidentiality.

In the second phase, executed on demand by consumers, a consumer requests a zero-knowledge proof (ZKP) of non-membership with respect to specific policy constraints (e.g., known CVEs, restricted licenses, or organizationally prohibited components). The consumer verifies the proof locally, ensuring both its correctness and its consistency with the on-chain commitment and the specified policy constraints.

The architecture is guided by the following design principles:
\begin{enumerate}
    \item Decoupling: The architecture separates software build and commitment from compliance verification. The provider captures the software state once, enabling subsequent queries related to newly disclosed vulnerabilities or evolving license constraints to be resolved without rebuilding artifacts. This supports individualized assurances for consumers with distinct policy requirements.
    \item Confidentiality: The provider can demonstrate the absence of specific policy violations without disclosing the full SBOM or the complete dependency graph.
    \item On-demand Efficiency: Proofs are generated only when required, ensuring that computational resources are utilized only when verification is required.
    \item Distributed Disclosure Control: Each consumer receives a proof tailored to their specific policy constraints. As a result, disclosed information is partitioned across independent consumers, preventing any single party from reconstructing the full dependency graph.
    \item Integration Simplicity: The commitment and proving infrastructure are established once per organization. The CI/CD workflow remains generic and reusable across projects, thereby minimizing integration effort and adoption overhead.
\end{enumerate}

\subsection{Threat Model}\label{sec:threat-model}
We consider a threat model in which the software provider acts as a potentially malicious prover that may attempt to generate invalid proofs to conceal the presence of policy-violating dependencies. Consumers are modeled as honest verifiers who correctly follow the verification protocol and seek only to validate compliance with specified policy constraints, without attempting to infer additional information about the SBOM.
The adversary is assumed to have full control over the provider-side infrastructure, including the proving system and request handling logic, but cannot break standard cryptographic assumptions or tamper with the underlying blockchain.

We make the following assumptions:
\begin{itemize}
    \item Blockchain Integrity: The public blockchain provides an immutable and publicly auditable ledger. Once commitments are recorded, they cannot be altered or removed. 
    \item Cryptographic Primitives: The employed hash function (SHA-256) is collision-resistant, and the zk-STARK construction provides computational soundness.
    \item CI Pipeline Integrity: The provider’s CI pipeline correctly transforms the build artifact into its corresponding SMT representation. This assumption is supported through periodic audits (\Cref{sec:ci}).
\end{itemize}

Under these assumptions, the system satisfies the standard properties of zero-knowledge proofs:
(1) \emph{Completeness}: An honest provider can generate a valid proof that correctly attests compliance with the consumer’s policy constraints.
(2) \emph{Soundness}: A malicious provider cannot produce a valid proof asserting compliance $\texttt{Compliant} = \texttt{True}$ if the SBOM contains dependencies that violate the specified policy constraints. 
(3) \emph{Zero-Knowledge}: The proof reveals only the compliance outcome with respect to the queried policy constraints and does not disclose any additional information about the SBOM.

In addition, \textsc{TrustBOM} provides \emph{non-repudiation}, as blockchain-based commitments prevent providers from retroactively altering or denying prior claims.

The model does not protect against information leakage through repeated queries or side-channel observations, which are addressed as deployment considerations in (\cref{sec:limits}).

\section{Implementation}\label{sec:implementation}

This section describes the implementation of TrustBOM across the provider’s CI pipeline, the proving system, and the consumer-side verification process. The implementation\footnote{\scriptsize\url{https://github.com/tuberlin-blockchain-prototyping}} and documentation\footnote{\scriptsize\url{https://tuberlin-blockchain-prototyping.github.io/trustbom-docs}} are publicly available and open-source.

\subsection{Provider CI Pipeline}\label{sec:ci}
\begin{figure*}[t]
    \centering
    \includegraphics[width=\linewidth]{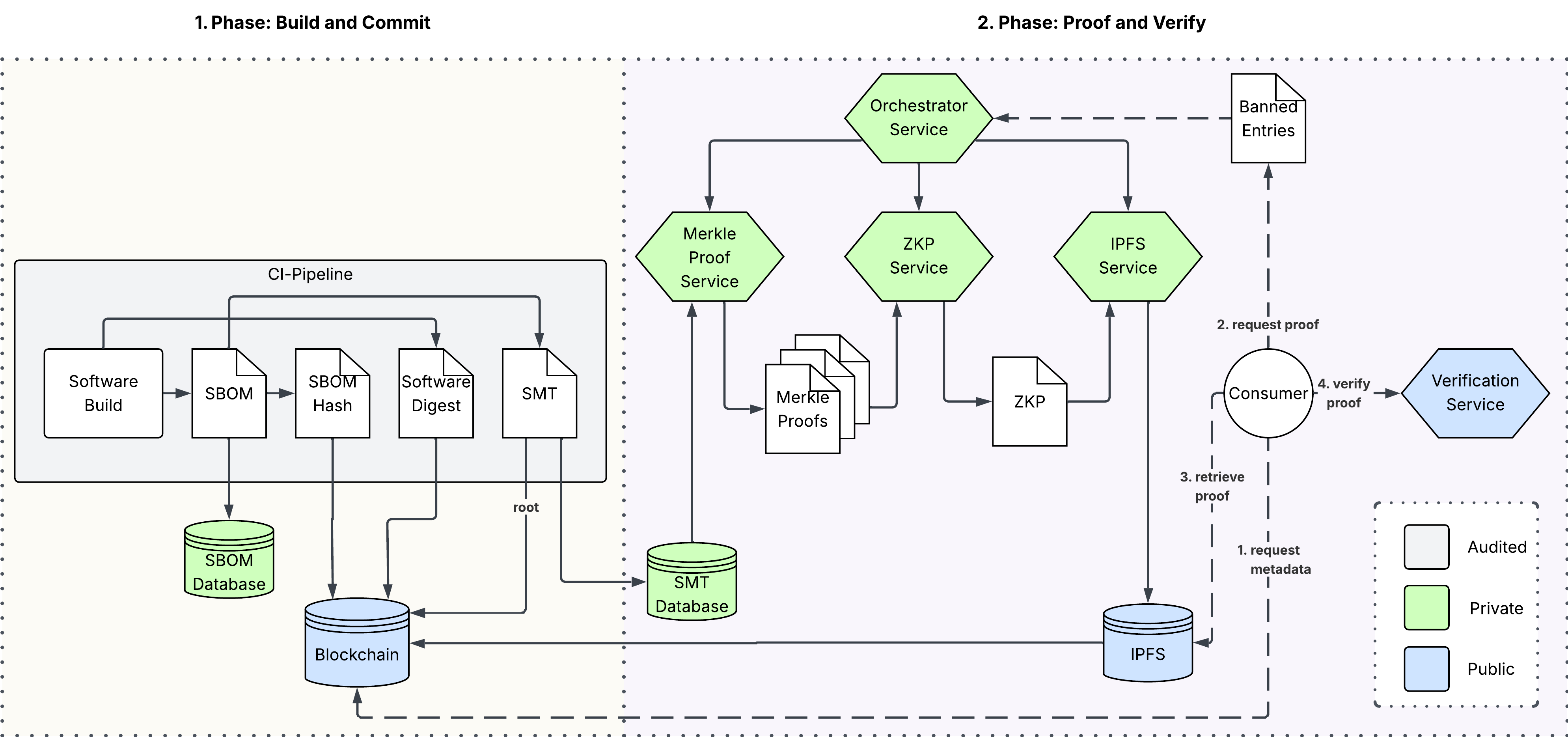}
    \caption{Detailed TrustBOM Architecture $\vert$ Dashed arrows for consumer initiated requests. All other arrows are for automated interactions}
    \label{fig:architecture}
\end{figure*}

The commitment phase is executed within the provider’s CI pipeline (\cref{fig:architecture}) and establishes the cryptographic foundation for subsequent proofs. For each release, the pipeline builds the software artifact and generates an SBOM using CycloneDX-compatible tooling. Build provenance is attested using SLSA\footnote{\scriptsize\url{https://slsa.dev/}, [Accessed: 26 Jan. 2026]}, binding the SBOM to the corresponding build process.

To enable efficient non-membership proofs, the SBOM is deterministically transformed into a Sparse Merkle Tree (SMT) of depth 256 using a versioned, open-source CLI tool. Each dependency is represented by its PURL, hashed via SHA-256 to determine the leaf index. Presence is encoded by setting the corresponding leaf to 1, while all other leaves remain 0. This mapping supports fixed-size indexing for variable-length identifiers and accommodates extended identifiers such as dependency–license combinations.

The cleartext SBOM and SMT are stored off-chain within the provider’s infrastructure. To bind these artifacts, the CI pipeline computes hashes of the source code, container image, and SBOM, and records them together with the SMT root on a public blockchain. This creates an immutable mapping between the software artifact and its committed dependency structure.

Since the SBOM-to-SMT transformation is performed off-chain, correctness relies on the CI Pipeline Integrity assumption (\cref{sec:threat-model}), which is validated through periodic audits. These audits verify that the SMT accurately reflects the built artifact and are required only per pipeline version rather than per proof.

\subsection{Software Provider Proving System}\label{sec:proving-system}
Once the software state is committed on-chain, the proving system handles consumer requests for compliance verification (\cref{fig:architecture}). The system is implemented as a microservice architecture hosted by the provider, ensuring that sensitive data remains internal.

Upon receiving a request containing an SMT root and a set of policy constraints (e.g., PURLs or license identifiers), the workflow proceeds as follows:

\begin{enumerate}
    \item The \texttt{Orchestrator Service} retrieves the corresponding \gls{smt} and invokes the \texttt{Merkle Proof Service}.
    \item The \texttt{Merkle Proof Service} generates Merkle non-membership proofs for each constraint. Proofs are compressed using a bitmap that encodes empty subtrees, allowing default hashes to replace omitted siblings and reducing data size.    
    \item The \texttt{Orchestrator} submits the SMT root (public input) and compressed proofs (private inputs) to the \texttt{Proving Service}, which executes verification logic within a RISC Zero zkVM and generates a zero-knowledge proof.
    \item Finally, the resulting proof artifact is stored via an \texttt{IPFS Service}, and its content identifier is anchored on-chain to provide a tamper-evident audit record.    
\end{enumerate}

\input{sections/04_algorithm}

\subsubsection{Algorithm Overview:} The \cref{alg:zkvm_logic} iterates over each non-membership proof, performing three checks: (1) the leaf value is zero (dependency absent), (2) the leaf index matches the SHA-256 hash of the queried PURL, and (3) the Merkle path reconstructs to the committed root. The bitmap compression allows skipping empty subtrees by substituting precomputed default hashes.

\subsubsection{ZK Circuit Logic:}
The zkVM guest code (Algorithm 1) verifies each non-membership proof by checking (i) absence at the leaf, (ii) correctness of the leaf index derived from the PURL, and (iii) consistency of the reconstructed Merkle path with the committed root. Bitmap-based compression reduces proof size by omitting empty subtrees.
The circuit aggregates all verified constraints and commits both the SMT root and a hash of the queried constraints to the public output. This ensures that the generated proof attests to the exact policy constraints requested, while enabling efficient batch verification within a single proof.

This logic ensures that a valid proof serves as a guarantee that the exact policy constraints requested were checked.
Consequently, this allows our system to create a \gls{zkp} over the correct verification of Merkle proofs and still provide the same assurances as iteratively proving non-membership of policy constraints in a \gls{sbom}, effectively shifting complex logic outside the \gls{zkvm} without sacrificing any trust guarantees. The use of RISC Zero's pre-compiles for SHA-256 accelerates the verification further, since the majority of computation tasks during the \gls{zkp} generation are hash operations.

\subsection{Consumer Verification}\label{sec:consumer-verification}
Consumer-side verification is performed locally to avoid reliance on the provider. The verification process consists of three steps:

\begin{enumerate}
    \item The consumer uses a RISC Zero verifier to validate the proof and confirm it was generated by the expected guest code via its 
    ImageID\footnote{\scriptsize\url{https://dev.risczero.com/terminology\#image-id}, [Accessed: 2 Feb. 2026]}.
    \item The \texttt{SMTRoot} included in the proof is checked against the corresponding blockchain commitment.  
    \item The consumer recomputes the hash of the requested policy constraints and verifies it matches the value included in the proof output.
\end{enumerate}

Together, these checks ensure that the software satisfies the specified policy constraints without revealing the underlying SBOM.

\vspace{1em}
\noindent
In summary, TrustBOM enables confidentiality-preserving compliance verification by decoupling artifact commitment from proof generation. Consumers obtain cryptographic assurances tailored to their policy constraints, while providers retain full control over their dependency information.

%% file: sections/04_algorithm.tex
\begin{algorithm}[H]
\scriptsize
\SetAlgoLined
\DontPrintSemicolon
\KwIn{Proofs $\mathcal{P}$, PublicInput $Root_{in}$}
\KwOut{Commitment $\{Root_{out}, H_{banned}, Compliant\}$}

$H_{banned} \leftarrow \text{SHA256}(\{p.purl \mid p \in \mathcal{P}\})$\;

\ForEach{$proof \in \mathcal{P}$}{
    \If{$proof.value \neq "0"$}{
        \Return{Commit($Root_{in}, H_{banned}, \text{False}$)}
    }
    
    $LeafIndex \leftarrow \text{ParseHex}(proof.leaf\_index)$\;
    
    \If{$LeafIndex \neq \text{SHA256}(proof.purl)$}{
        \Return{Commit($Root_{in}, H_{banned}, \text{False}$)}
    }

    $Current \leftarrow \text{Hash}(proof.value)$\;
    $Bitmap \leftarrow \text{ParseHex}(proof.bitmap)$\;
    $SibPtr \leftarrow 0$\;

    \For{$d \leftarrow 0$ \KwTo $255$}{
        \eIf{\text{BitAt}($Bitmap, d$) == 1}{
            $Sibling \leftarrow \text{ParseHex}(proof.siblings[SibPtr])$\;
            $SibPtr \leftarrow SibPtr + 1$\;
        }{
            $Sibling \leftarrow DEFAULTS[d]$
        }
        
        \eIf{\text{BitAt}($LeafIndex, d$) == 0}{
            $Current \leftarrow \text{Hash}(Current, Sibling)$\;
        }{
            $Current \leftarrow \text{Hash}(Sibling, Current)$\;
        }
    }
    
    \If{$Current \neq Root_{in}$}{
        \Return{Commit($Root_{in}, H_{banned}, \text{False}$)}
    }
}

\Return{Commit($Root_{in}, H_{banned}, \text{True}$)}\;
\caption{Compressed Non-Membership Verification}
\label{alg:zkvm_logic}
\end{algorithm}

%% file: sections/05_evaluation.tex
\section{Evaluation}\label{sec:eval}
TrustBOM relies on zero-knowledge proofs to provide confidentiality-preserving SBOM-based compliance; however, such proofs introduce computational overhead. This evaluation examines whether TrustBOM can support enterprise-scale software delivery by analyzing the scalability and cost of proof generation. Specifically, we investigate whether proof generation scales predictably with the number of policy constraints, whether latency remains compatible with enterprise workflows, and whether the associated costs are economically feasible for real-world deployment.

\subsection{Methodology}
In TrustBOM, a proof constitutes a cryptographic attestation that a given SBOM satisfies a set of consumer-defined policy constraints without revealing its contents. The computational cost of proof generation increases with the number of constraints to be verified.

Preliminary benchmarking identifies the \texttt{Proving Service} as the dominant contributor to overall resource consumption, with all other system components incurring negligible overhead. Consequently, the evaluation focuses on proof generation performance, analyzing its behavior as the number of policy constraints increases.

We measure three primary metrics as functions of policy constraint count (the number of policy constraints verified per proof request, serving as the independent variable): proving time (the total time required to generate a verifiable proof), proof size (the file size of the generated \gls{zkp}), and computational cycles (the number of zkVM execution cycles). 

Proving time directly affects the latency experienced by consumers waiting for attestation results, and therefore determines whether TrustBOM can fit within the time constraints of enterprise software delivery pipelines. Proof size affects transmission and storage costs relevant for providers. Computational cycles serve as a hardware-agnostic measure of circuit complexity, reflecting the underlying cost of the zkVM execution independent of the machine it runs on.

This setup isolates proof generation as the dominant cost factor, allowing us to evaluate the scalability of the core cryptographic mechanism independent of auxiliary system components.

\subsection{Experiment Setup}
 Table~\ref{tab:experimental_setup} summarizes the experimental configuration. The hardware utilized has GPU acceleration and is accessible on all major cloud platforms.

\begin{table}[htbp]
\caption{Experimental Configuration}
\label{tab:experimental_setup}
\centering
\resizebox{0.6\columnwidth}{!}{
\begin{tabular}{ll}
\hline
\textbf{Component} & \textbf{Specification} \\
\hline
Processor & AMD EPYC 8224P (24-Core, 48 threads) \\
GPU & NVIDIA L4 (23 GB VRAM) \\
Memory & 188 GB RAM \\
Operating System & Ubuntu 22.04.5 LTS \\
RISC Zero zkVM & Version 3.0 with CUDA acceleration \\
CUDA Version & 12.8 \\
\hline
\end{tabular}
}
\end{table}

We use a real-world \gls{sbom} generated from a production application from German software company adesso SE\footnote{\scriptsize\url{https://www.adesso.de/en/}, [Accessed: 2 Feb. 2026]} in CycloneDX format. The policy constraints comprise 200 vulnerable Maven packages, including Log4Shell-affected versions (log4j-core, log4j-api), Jackson databind, Apache Commons, Spring Framework, Struts, and Tomcat. Note that the specific PURLs used do not affect performance.

\subsection{Experiment Protocol}
The benchmark measures \gls{zkp} generation performance as a function of the number of non-membership proofs verified. Proofs are generated across seven experiments workloads of increasing constraint count: $N \in \{2, 5, 10, 20, 50, 100, 200\}$. For each experiment, the first N constraints are drawn from the reference set of 200, so that smaller experiments are strict subsets of larger ones. This ensures comparability across runs. Merkle non-membership proofs are pre-computed for all 200 constraints against the SBOM's sparse Merkle tree before benchmarking begins, isolating zero-knowledge proof generation as the variable under measurement.

The \texttt{Proving Service} receives a batch of $N$ Merkle non-membership proofs along with the tree root and generates a single \gls{zkp} attesting to the validity of all proofs. Each proof demonstrates that a specific policy constraint (package identifier) does not exist in the \gls{sbom}'s Merkle tree. The \texttt{Proving Service} executes the verification logic and produces a cryptographic receipt that can be independently verified.

For each workload $N$, measurements are repeated ten times to account for variance. This yields 70 experimental runs (7 workload $\times$ 10 repetitions). We report mean values with min-max ranges for all three metrics.

\subsection{Results}

\begin{figure}[htbp]
    \centering
    \includegraphics[width=\columnwidth]{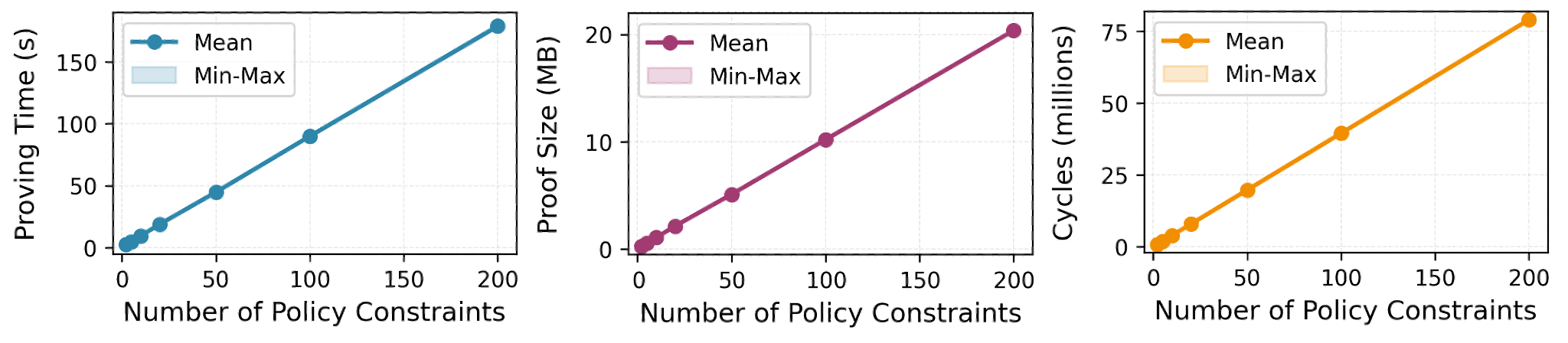}
    \caption{Proof generation performance relative to policy constraint count, across 10 experiment runs. \textit{Note}: error bands are present but largely obscured by the mean line due to low variance}
    \label{fig:benchmark}
\end{figure}

\Cref{fig:benchmark} presents the benchmark results across three metrics: proving time, proof size, and computational cycles.

All three metrics exhibit linear scaling with the number of policy constraints verified. Proving time increases from 2.4 seconds for 2 policy constraints to 178.9 seconds for 200 policy constraints, yielding approximately 0.9 seconds per policy constraint. Proof size scales from 0.28~MB to 21.4~MB, corresponding to roughly 0.1~MB per policy constraint. Computational cycles range from 1.0 million to 78.9 million, maintaining a consistent ratio of approximately 0.4 million cycles per policy constraint. The tight min-max bands visible in the figure indicate low variance across runs (standard deviation below 1\%), demonstrating highly reproducible performance.

The linear scaling demonstrates that the \texttt{Proving Service} performs constant work per Merkle non-membership proof verification. This predictable relationship enables resource estimation for arbitrary policy constraint counts: verifying $N$ constraints requires approximately $0.9N$ seconds and produces a proof of $0.1N$~MB. Utilizing Merkle Proofs also ensures that proving time is independent from the number of entries in the original \gls{sbom}.
This confirms that the computational cost of TrustBOM is determined by the number of policy constraints rather than the size or complexity of the underlying SBOM, validating the architectural design principle of constraint-linear verification.

These results confirm that TrustBOM is practical for production deployment. Verifying 200 policy constraints completes in under three minutes on an NVIDIA L4, available at approximately \$0.70/hour\footnote{\scriptsize\url{https://modal.com/blog/nvidia-l4-price-article}, [Accessed: 2 Feb. 2026]} on major cloud platforms. The linear scaling ensures that larger policy constraint sets remain tractable, and the low variance guarantees predictable execution times for integration into automated pipelines.

In contrast to approaches that require full SBOM disclosure or repeated vulnerability scanning, TrustBOM enables targeted verification with predictable, constraint-linear cost.

\subsection{Discussion}
To contextualize these results against real-world vulnerability databases, we estimate proof generation times for full ecosystem coverage using data from Google’s OSV.dev\footnote{\scriptsize\url{https://osv.dev/list}, [Accessed: 26 Jan. 2026]}. For the Maven ecosystem, which includes 6,125 known vulnerable packages, proving non-membership for all entries would require approximately 92 minutes at an estimated cost of \$1.07. For PyPI, with 17,594 vulnerabilities, proof generation would complete in approximately 4.4 hours (\$3.08). Even for the npm ecosystem, which includes 214,233 vulnerabilities, full coverage remains tractable at approximately 53 hours (\$37.49).

These results represent a worst-case upper bound in which proof generation covers all known vulnerabilities within an entire ecosystem. In practice, enterprise policies are typically defined over specific subsets of constraints, such as critical vulnerabilities, restricted licenses, or organizationally prohibited components, resulting in substantially lower proof generation times. Moreover, due to the decoupled architecture of TrustBOM, proof generation is performed independently of the software build process and can be executed on demand or within external verification workflows. As a result, proof generation does not impact build latency and can be scheduled according to application-specific performance requirements.

Overall, these estimates indicate that comprehensive vulnerability attestation across entire ecosystems is economically feasible, while typical usage scenarios incur significantly lower computational cost. Organizations with stricter latency requirements can further reduce proving time by leveraging high-performance hardware, such as NVIDIA A100 or H100 GPUs.

Blockchain transaction costs are negligible compared to proof generation costs. The smart contract stores only constant-size commitments per transaction, independent of \gls{sbom} size. Under current market conditions (ETH: \$3,081; gas price: 0.051 gwei)\footnote{\scriptsize\url{https://etherscan.io/},  [Accessed: 14 Dec. 2025]}, registering an \gls{smt} root costs approximately \$0.03, while recording a compliance proof costs \$0.04, resulting in a total cost of \$0.07 per complete verification cycle on Ethereum Layer 1.

These costs remain constant regardless of whether the \gls{smt} contains tens or thousands of dependencies. Furthermore, in light of the blockchain trilemma balancing decentralization, security, and scalability, organizations can further reduce transaction costs by deploying the system on alternative blockchains or Layer 2 solutions optimized for lower fees.

This demonstrates that TrustBOM supports both interactive verification for targeted policy checks and batch-oriented compliance workflows at ecosystem scale.

%% file: sections/06_conclusion.tex
\section{Conclusion}\label{sec:conclusion}
This paper introduced TrustBOM, a confidentiality-preserving SBOM architecture for verifiable compliance across organizations. It enables software providers to attest to the absence of specific vulnerabilities or restricted licenses for different consumers, while fully concealing proprietary dependency graphs. By decoupling disclosure from assurance, TrustBOM transforms compliance verification into a constraint-linear process whose cost depends on the number of policy constraints rather than the size of the SBOM. The resulting commitments serve as durable, tamper-evident audit artifacts aligned with enterprise platform engineering and long-term governance requirements.

Our evaluation demonstrates the practical feasibility of this approach in production environments. Proof generation scales linearly at approximately 0.9 seconds per policy constraint on commodity hardware (NVIDIA L4), with predictable resource consumption that supports capacity planning in DevSecOps workflows. Even ecosystem-wide vulnerability attestation remains economically feasible: verifying the absence of all known Maven vulnerabilities, for example, incurs a cost of approximately \$1.07.

These results indicate that TrustBOM provides a cost-efficient and scalable mechanism for confidentiality-preserving SBOMs with verifiable compliance across software supply chains, addressing key barriers that have so far limited the adoption of SBOMs in cross-organizational settings. By decoupling disclosure from verification, TrustBOM demonstrates that strong confidentiality guarantees and practical supply chain assurance are not inherently conflicting, but can be achieved simultaneously within enterprise architectures.

\subsection{Limitations}\label{sec:limits}
Several limitations warrant discussion. First, the proposed approach inherits the fundamental limitations of SBOMs themselves. The system assumes a complete and accurate SBOM representation of the software; however, if SBOM generation tools fail to capture transitive dependencies or misidentify component versions \cite{Yu_Song_SBOMCorrectness_2024}, the resulting guarantees apply only to this incomplete or incorrect representation. Improving SBOM fidelity remains an orthogonal challenge that is not addressed by our architecture.

Second, the security guarantees depend on the correctness of the underlying cryptographic implementation. While we rely on RISC Zero, a mature and actively audited platform, vulnerabilities in the \gls{zkvm} can compromise soundness properties. For example, CVE-2025-61588\footnote{\scriptsize\url{https://nvd.nist.gov/vuln/detail/CVE-2025-61588}, [Accessed: 12 Feb. 2026]} exposed a critical flaw in RISC Zero's \texttt{sys\_read} system call that allowed arbitrary code execution within the guest, enabling malicious hosts to forge proofs for invalid computations. Such vulnerabilities underscore that cryptographic guarantees are contingent on implementation correctness; production deployments must track security advisories and maintain updated dependencies.

Third, the system cannot verify that the provided software corresponds to the software committed in the \gls{smt}. A malicious provider could generate a valid commitment, pass all compliance checks, and subsequently deploy different software. While periodic CI pipeline audits mitigate this risk, they cannot provide continuous assurance. Even with auditor access, verifying runtime deployment fidelity in cloud environments remains challenging.

Finally, the proof-request interface introduces potential information leakage vectors. An adversarial consumer could submit excessive proof requests to incrementally reconstruct the \gls{sbom} through repeated non-membership queries. Production deployments may implement rate limiting, request authentication, anomaly detection, or a human-in-the-loop to prevent such reverse-engineering attacks. Additionally, proof responses should incorporate configurable delays to prevent timing-based attacks that could exploit response latency variations.

\subsection{Future Work}

Several directions merit further investigation. First, extending the system to support membership proofs would enable positive attestations such as license compliance verification. 
This can be achieved using the same \gls{smt} representation and Merkle proof construction, requiring only modifications to the constraint logic in the proving service.

Second, recursive proof aggregation\footnote{\scriptsize\url{https://risczero.com/blog/proof-composition}, [Accessed: 26 Jan. 2026]} could enable compliance dashboards that verify multiple releases with a single proof, reducing verification overhead. 

Third, bridging the gap between committed artifacts and runtime deployment could leverage confidential computing. If providers deploy within trusted execution environments (TEEs) supporting remote attestation, consumers could obtain hardware-signed evidence of which binary is running. Aligning TEE measurements with the artifact hashes committed on-chain would require careful design of the deployment pipeline, but could extend cryptographic assurance from CI commitment through to runtime.

More broadly, integrating build-time integrity guarantees with post-build compliance verification represents a promising direction. Approaches based on deterministic builds and TEE-based attestation can provide verifiable evidence of artifact provenance within CI pipelines \cite{castillo_ciprotocol_2026}, while TrustBOM enables confidential\-ity-preserving verification of software composition after artifact generation. Combining these techniques could enable end-to-end verifiable supply chains spanning build, distribution, and deployment. 

Finally, extending the architecture to multi-provider supply chains (beyond the single-provider setting considered in this work), where sub-providers also issue cryptographic commitments for their components, would enable end-to-end supply chain verification while preserving confidentiality at each tier. Frameworks such as the Trusted Compute Unit (TCU)~\cite{11114627} provide a foundation for this setting, supporting interoperable proofs across heterogeneous technologies (TEEs and zkVMs) with blockchain-based traceability.

As regulatory frameworks increasingly mandate software supply chain accountability, architectures that balance compliance requirements with confidentiality concerns are becoming essential. TrustBOM demonstrates that zero-know\-ledge proofs provide a practical foundation for this balance, enabling verifiable supply chain assurance without compromising proprietary information.

%% file: references.bib
@article{Williams_ResearchDirections_2025, title={Research Directions in Software Supply Chain Security}, volume={34}, ISSN={1049-331X}, DOI={10.1145/3714464}, abstractNote={Reusable software libraries, frameworks, and components, such as those provided by open source ecosystems and third-party suppliers, accelerate digital innovation. However, recent years have shown almost exponential growth in attackers leveraging these software artifacts to launch software supply chain attacks. Past well-known software supply chain attacks include the SolarWinds, log4j, and xz utils incidents. Supply chain attacks are considered to have three major attack vectors: through vulnerabilities and malware accidentally or intentionally injected into open source and third-party dependencies/components/containers; by infiltrating the build infrastructure during the build and deployment processes; and through targeted techniques aimed at the humans involved in software development, such as through social engineering. Plummeting trust in the software supply chain could decelerate digital innovation if the software industry reduces its use of open source and third-party artifacts to reduce risks. This article contains perspectives and knowledge obtained from intentional outreach with practitioners to understand their practical challenges and from extensive research efforts. We then provide an overview of current research efforts to secure the software supply chain. Finally, we propose a future research agenda to close software supply chain attack vectors and support the software industry.}, number={5}, journal={ACM Trans. Softw. Eng. Methodol.}, author={Williams, Laurie and Benedetti, Giacomo and Hamer, Sivana and Paramitha, Ranindya and Rahman, Imranur and Tamanna, Mahzabin and Tystahl, Greg and Zahan, Nusrat and Morrison, Patrick and Acar, Yasemin and Cukier, Michel and Kästner, Christian and Kapravelos, Alexandros and Wermke, Dominik and Enck, William}, year={2025}, month=may, pages={146:1-146:38} }

@techreport{ntia2021sbom,
  author = {{National Telecommunications and Information Administration}},
  title = {The Minimum Elements For a Software Bill of Materials (SBOM)},
  institution = {U.S. Department of Commerce},
  year = {2021},
  month = {July},
  url = {https://www.ntia.gov/files/ntia/publications/sbom_minimum_elements_report.pdf}
}

@inproceedings{Xia_SBOM_2023,
    title = {An Empirical Study on Software Bill of Materials: Where We Stand and the Road Ahead},
    author = {Xia, Boming and Bi, Tingting and Xing, Zhenchang and Lu, Qinghua and Zhu, Liming},
    booktitle = {2023 IEEE/ACM 45th International Conference on Software Engineering (ICSE)},
    pages = {2630--2642},
    year = {2023},
    doi = {10.1109/ICSE48619.2023.00219}
}

@inproceedings{Xia_blockchain_2024,
    author = {Xia, Boming and Zhang, Dawen and Liu, Yue and Lu, Qinghua and Xing, Zhenchang and Zhu, Liming},
    title = {Trust in Software Supply Chains: Blockchain-Enabled SBOM and the AIBOM Future},
    year = {2024},
    isbn = {9798400705656},
    publisher = {Association for Computing Machinery},
    address = {New York, NY, USA},
    doi = {10.1145/3643662.3643957},
    booktitle = {Proceedings of the 2024 ACM/IEEE 4th International Workshop on Engineering and Cybersecurity of Critical Systems (EnCyCriS) and 2024 IEEE/ACM Second International Workshop on Software Vulnerability},
    pages = {12--19},
    numpages = {8},
    location = {Lisbon, Portugal},
    series = {EnCyCriS/SVM '24}
}

@inproceedings{Yu_Song_SBOMCorrectness_2024, title={On the Correctness of Metadata-Based SBOM Generation: A Differential Analysis Approach}, DOI={10.1109/DSN58291.2024.00018}, abstractNote={Amidst rising concerns of software supply chain attacks, the Software Bill of Materials (SBOM) has emerged as a pivotal tool, offering a detailed listing of software components to manage vulnerabilities, dependencies, and licensing. While many SBOM generation tools are extensively used in both commercial and open-source realms, the correctness of these tools remains largely unscrutinized. To date, there has not been a systematic study addressing the correctness of contemporary SBOM generation solutions. In this paper, we conduct a large-scale differential analysis of the correctness of four popular SBOM generators. Surprisingly, our evaluation reveals all four SBOM generators exhibit inconsistent SBOMs and dependency omissions, leading to incomplete and potentially inaccurate SBOMs. Moreover, we construct a parser confusion attack against these tools, introducing a new attack vector to conceal malicious, vulnerable, or illegal packages within the software supply chain. Drawing from our analysis, we propose best practices for SBOM generation and introduce a benchmark to steer the development of more robust SBOM generators.},  booktitle={2024 54th Annual IEEE/IFIP International Conference on Dependable Systems and Networks (DSN)}, author={Yu, Sheng and Song, Wei and Hu, Xunchao and Yin, Heng}, year={2024}, pages={29–36} }

@article{doi:10.1137/0218012,
author = {Goldwasser, Shafi and Micali, Silvio and Rackoff, Charles},
title = {The Knowledge Complexity of Interactive Proof Systems},
journal = {SIAM Journal on Computing},
volume = {18},
number = {1},
pages = {186-208},
year = {1989},
doi = {10.1137/0218012}
}

@techreport{ntia_sbomdefinition_2019,
author = {Framing Working Group},
title={Framing Software Component Transparency: Establishing a Common Software Bill of Material (SBOM)},
institution = {NTIA},
  year        = {2019},
  note = {\url{https://www.ntia.gov/files/ntia/publications/framingsbom_20191112.pdf}}
}

@InProceedings{10.1007/978-3-319-47560-8_13,
author="Dahlberg, Rasmus
and Pulls, Tobias
and Peeters, Roel",
editor="Brumley, Billy Bob
and R{\"o}ning, Juha",
title="Efficient Sparse Merkle Trees",
booktitle="Secure IT Systems",
year="2016",
publisher="Springer International Publishing",
address="Cham",
pages="199--215",
isbn="978-3-319-47560-8"
}

@INPROCEEDINGS{8726497,

  author={Eberhardt, Jacob and Tai, Stefan},

  booktitle={2018 IEEE International Conference on Internet of Things (iThings) and IEEE Green Computing and Communications (GreenCom) and IEEE Cyber, Physical and Social Computing (CPSCom) and IEEE Smart Data (SmartData)}, 

  title={ZoKrates - Scalable Privacy-Preserving Off-Chain Computations}, 

  year={2018},

  volume={},

  number={},

  pages={1084-1091},

  doi={10.1109/Cybermatics_2018.2018.00199}}

@INPROCEEDINGS{10936765,

  author={Song, Ahyun and Seo, Euiseong and Kim, Heeyoul},

  booktitle={2025 27th International Conference on Advanced Communications Technology (ICACT)}, 

  title={BC-SBOM: Blockchain-based SBOM Management System}, 

  year={2025},

  volume={},

  number={},

  pages={169-174},

  doi={10.23919/ICACT63878.2025.10936765}}

@INPROCEEDINGS{9343074,

  author={Mukta, Rahma and Martens, James and Paik, Hye-young and Lu, Qinghua and Kanhere, Salil S.},

  booktitle={2020 IEEE 19th International Conference on Trust, Security and Privacy in Computing and Communications (TrustCom)}, 

  title={Blockchain-Based Verifiable Credential Sharing with Selective Disclosure}, 

  year={2020},

  volume={},

  number={},

  pages={959-966},

  doi={10.1109/TrustCom50675.2020.00128}
}

@INPROCEEDINGS{11114627,

  author={Castillo, Fernando and Heiss, Jonathan and Werner, Sebastian and Tai, Stefan},

  booktitle={2025 IEEE International Conference on Blockchain and Cryptocurrency (ICBC)}, 

  title={Trusted Compute Units: A Framework for Chained Verifiable Computations}, 

  year={2025},

  volume={},

  number={},

  pages={1-9},

  doi={10.1109/ICBC64466.2025.11114627}}

@inproceedings{castillo_ciprotocol_2026,
  title={An Evidence-driven Protocol for Trustworthy CI Pipelines},
  author={Castillo, Fernando and Brito, Eduardo and Pullonen-Raudvere, Pille and Werner, Sebastian and Tai, Stefan},
  booktitle={30th International Conference on Enterprise Design, Operations, and Computing (EDOC)},
  year={2026}
}
